\documentclass[%
superscriptaddress,
preprintnumbers,
 amsmath,amssymb,
 aps,
]{revtex4-2}

\usepackage{graphicx}
\usepackage{dcolumn}
\usepackage{bm}
\usepackage{slashed}
\usepackage{hyperref}
\usepackage{xcolor}

\DeclareMathOperator{\diag}{diag}
\newcommand{\hx}[1]{#1}
\begin{document}


\title{Ultralight dark matter in long-baseline accelerator neutrino oscillations}

\author{Xin-Qiang Li}%
\email{xqli@mail.ccnu.edu.cn}
\affiliation{Institute of Particle Physics and Key Laboratory of Quark and Lepton Physics~(MOE), Central China Normal University, Wuhan, Hubei 430079, China}

\author{Hai-Xing Lin}
\email{linhx55@mails.ccnu.edu.cn}
\affiliation{Institute of Particle Physics and Key Laboratory of Quark and Lepton Physics~(MOE), Central China Normal University, Wuhan, Hubei 430079, China}
\affiliation{School of Physics, Sun Yat-Sen University, 510275 Guangzhou, China}

\author{Jian Tang}
\email{tangjian5@mail.sysu.edu.cn}
\affiliation{School of Physics, Sun Yat-Sen University, 510275 Guangzhou, China}

\author{Sampsa Vihonen}
\email{vihonen@kth.se}
\affiliation{Department of Physics, School of Engineering Sciences,
KTH Royal Institute of Technology and The Oskar Klein Centre, AlbaNova University Center,
Roslagstullsbacken 21, SE--106 91 Stockholm, Sweden}
\affiliation{School of Physics, Sun Yat-Sen University, 510275 Guangzhou, China}


\begin{abstract}
We present a systematic study of the effects of ultralight dark matter (ULDM) on neutrino oscillations using the latest long-baseline data from the T2K and NO$\nu$A experiments. Our analysis covers both flavor-universal and flavor-general scalar interactions, as well as vector interactions associated with the $L_e - L_\mu$ and $L_\mu - L_\tau$ gauge symmetries. Importantly, we explicitly consider the coherence properties of the ULDM by incorporating the resulting stochastic fluctuations into our statistical analysis. We find that in the low-mass regime, $m_\phi \lesssim 10^{-17}$~eV, where the stochastic effects are maximal, the constraints on the ULDM couplings are relaxed by roughly an order of magnitude compared to those in the high-mass regime, $m_\phi \gtrsim 10^{-15}$~eV, where such fluctuations are effectively averaged out. While the combined T2K and NO$\nu$A datasets impose nontrivial exclusion limits on the ULDM interactions, we do not find statistically significant evidence that these effects can alleviate the current tension in determining the charge-parity (CP) violating phase $\delta_{CP}$ between the two experiments. Therefore, it will be essential for future high-precision facilities to further probe the ULDM scenarios and achieve a definitive measurement of $\delta_{CP}$.
\end{abstract}


\maketitle

\section{Introduction}
\label{sec:intro}

Neutrino physics has now entered the era of precision measurements~\cite{Huber:2022lpm}. The recent release of observational datasets from the Tokai-to-Kamioka (T2K)~\cite{T2K:2023smv} and NuMI Off-axis $\nu_e$ Appearance (NO$\nu$A)~\cite{NOvA:2025tmb} experiments, culminating in a joint analysis presented in Ref.~\cite{T2K:2025wet}, signifies that the measured precision of some key oscillation parameters~---~most notably the atmospheric mass-squared splitting $\Delta m^2_{32}$ and the mixing angle $\theta_{23}$~---~has reached an unprecedented level, crossing the sub-$2\%$ threshold. These stringent experimental constraints not only provide a decisive foundation for addressing unresolved standard oscillation questions, such as the CP-violating phase $\delta_{CP}$ and the neutrino mass ordering, but also offer an exceptionally sensitive platform for probing subtle signals of physics beyond the Standard Model (SM). In particular, recent phenomenological analyses further exploit the T2K and NO$\nu$A datasets to probe dark-sector physics through neutrino oscillations~\cite{Alonso-Alvarez:2024wnh,Bai:2026kdq}, including the ultralight dark matter (ULDM) with a mass of about $m_\phi \sim$ 10$^{-22}$ eV~\cite{Brdar:2017kbt,Krnjaic:2017zlz}.

The nature of dark matter (DM), whose existence is supported by robust evidence of galactic rotation curves~\cite{Rubin:1980zd}, galaxy cluster movement~\cite{Bartelmann:1999yn,SDSS:2005xqv}, and cosmic microwave background (CMB) radiation~\cite{WMAP:2012nax,Planck:2018vyg}, remains one of the most compelling puzzles in modern physics. Although the weakly interacting massive particles have long been regarded as the leading candidates~\cite{Jungman:1995df}, their parameter space has been stringently constrained by the null results of direct detection experiments, such as LUX~\cite{LUX:2013afz} and PandaX-II~\cite{PandaX-II:2017hlx}. Furthermore, the conventional DM models have encountered problems at the small-scale level~\cite{Moore:1994yx}. In order to tackle these issues, numerous alternatives have been proposed, including the primordial black holes~\cite{Carr:2016drx} and the axions~\cite{Peccei:1977hh}, as well as the very light particles~\cite{Fayet:1980ad,Fayet:1980rr}. A good review on recent advances could be found, \textit{e.g.}, in Ref.~\cite{Oks:2021hef}.

The ULDM regime is a highly motivated topic within the DM community~\cite{Antypas:2022asj}. In addition to its implications in cosmology, the ultralight bosonic DM has been shown to aid with the questions of supermassive black holes~\cite{Davoudiasl:2021ijv}, pulsar timing arrays~\cite{Chowdhury:2023xvy}, St\"uckelberg axions~\cite{Coriano:2018uip}, and the age-velocity dispersion relation of the Milky Way disc~\cite{Chiang:2022rlx}. The ULDM could have further interesting effects in neutrino oscillation experiments, where even feeble couplings could lead to notable signatures in neutrino oscillations in terrestrial experiments~\cite{Krnjaic:2017zlz,Brdar:2017kbt,Liao:2018byh,Choi:2019ixb,Salla:2022dxc}. Such effects have been analyzed in various contexts, including an improved fit to the reactor neutrino data from KamLAND~\cite{Losada:2022uvr}, as well as the sensitivity projections for next-generation facilities, such as the DUNE and JUNO~\cite{Brdar:2017kbt,Krnjaic:2017zlz,Losada:2022uvr} and the proposed ESS$\nu$SB experiment~\cite{Cordero:2022fwb}. Beyond static matter-like effects, the ULDM interactions may induce observable temporal variations~\cite{Dev:2020kgz,Losada:2021bxx,Losada:2023zap}. The ULDM effect has also been considered in neutrino experiments in the case of pseudo-scalar~\cite{Huang:2018cwo,Brandenberger:2025gks} and vector fields~\cite{Brdar:2017kbt,Brzeminski:2022rkf,Gherghetta:2023myo,Arguelles:2023wvf,Chen:2023vkq}, which hold similarly promising prospects for future experiments.

Although extensive research has explored the ULDM constraints in neutrino experiments, certain physical nuances within the prevailing analytical frameworks warrant more rigorous scrutiny. First, due to its macroscopic de Broglie wavelength, the ULDM at local galactic densities can be approximated as a classically oscillating background field~\cite{Hu:2000ke,Hui:2016ltb}. Most existing studies model this field as a single-mode background with static amplitude and phase, and remove phase-dependent contributions in observables through time averaging. In this article, we argue that the realistic DM field is constrained by its finite coherence and should instead be regarded as a superposition of multiple near-degenerate frequency modes, which in turn leads to stochastic fluctuations in its effective amplitude and phase. When the experimental measurement or exposure time is significantly shorter than the DM coherence time, these fluctuations cannot be efficiently averaged out, and thus constitute a non-negligible physical effect~\cite{Centers:2019dyn,Foster:2017hbq}. Second, with respect to the construction of the scalar ULDM potential term, and assuming an independent (non-ULDM) origin for neutrino masses, the induced potential is inherently coupled to the intrinsic neutrino mass matrix~\cite{Ge:2018uhz,Berlin:2016woy}. Although this connection has been theoretically noticed, the extent to which the absolute neutrino mass scale $m_{\nu}$ directly influences the magnitude of the ULDM-induced potential~---~and consequently the robustness of the resulting experimental limits~---~has not been systematically scrutinized in previous phenomenological studies. This simplification potentially neglects a non-trivial physical dependence that extends beyond the standard oscillation parameters.

To address these scientific nuances, we perform in this work a systematic exclusion analysis of the ULDM parameter space across a broad mass range from $10^{-23}$ to $10^{-12}$~eV, utilizing the most recent joint datasets from T2K~\cite{T2K:2023smv} and NO$\nu$A~\cite{NOvA:2025tmb}. We aim to establish the most rigorous constraints to date by incorporating updated experimental observations. Crucially, our analysis explicitly incorporates the stochastic fluctuations of the DM field and examines their impact across different mass regimes, as determined by the interplay between the DM coherence time and the experimental exposure window. Operating within a joint likelihood framework, we extend our investigation across scalar and multiple vector-field models (including $L_e - L_\mu$ and $L_\mu - L_\tau$), and investigate how the absolute neutrino mass scale $m_{\nu}$ leads to a systematic shift in the exclusion limits for the scalar scenario.

This article is organized as follows. In section~\ref{sec:theory}, the theoretical formalism for neutrino interactions with the ULDMs is reviewed in light of the neutrino matter potentials. The datasets from the T2K and NO$\nu$A experiments and also the analysis methods are described in section~\ref{sec:data}. We present our phenomenological analysis of scalar and vector ULDMs in section~\ref{sec:results}. The conclusion made in section~\ref{sec:concl} synthesizes the results of the joint analysis and discusses future experimental prospects.

\section{Ultralight dark matter in neutrino oscillations}
\label{sec:theory}

In this section, we review the theoretical framework for the scalar and vector DM fields in the ultralight regime and discuss their phenomenological implications on the neutrino matter potentials in terrestrial experiments.

\subsection{Ultralight dark matter background}
\label{sec:theory1}

When the ULDM is present, neutrinos in terrestrial experiments can be seen to drift in a dark sea. The ULDM particles exhibit high occupation numbers and sufficiently low velocities to be treated as a classical coherent field~{\hx \cite{Berlin:2016woy}}, 
\begin{equation} \label{eq:phi_t}
    \phi(t) = \phi_0 \cos{(m_\phi t + \alpha)},
\end{equation}
where $m_\phi$ is the DM mass, $t$ denotes the laboratory time, and $\alpha$ is an arbitrary initial phase. Given the solar system's non-relativistic velocity $v \sim 10^{-3}$ relative to the Galactic Center, this field is dominated by the temporal term $m_\phi t$. We can therefore drop the spatial dependence. The DM field $\phi(t)$ can be regarded as a superposition of many field modes with randomly distributed phases. The resulting oscillation amplitude $\phi_0$ emerges from the interference among these modes and can be statistically understood as a random walk in the complex plane, leading to a Rayleigh distribution~\cite{Foster:2017hbq}
\begin{equation}
    p(\phi_0) = \frac{2\phi_0}{\phi_\mathrm{DM}^2}\exp\left(-\frac{\phi_0^2}{\phi_\mathrm{DM}^2}\right),
\end{equation}
where $\phi_\mathrm{DM} = \frac{\sqrt{2\rho_\mathrm{DM}}}{m_\phi}$ is determined by the local DM average density $\rho_{\rm DM} \simeq 0.3~\mathrm{GeV/cm^3}$~\cite{Nelson:2011sf,Arias:2012az}. The sampling number $N_\mathrm{DM}$ is determined by the ratio of the experimental exposure time $T_\mathrm{exp}$ to the coherence time $\tau_{c} \equiv {2\pi}/({m_\phi v^2})$~\cite{Schive:2014dra}, which characterizes the dephasing of the net field. 

When the running time of an experiment is $T_{\rm exp}\sim 10$ yr, the DM fields with mass $m_{\phi}\gtrsim 10^{-23}\;\mathrm{eV}$ complete many oscillation cycles during the data-taking period. Accordingly, phase averaging can be applied when computing event rates or oscillation probabilities, and this treatment will be adopted throughout our analysis. For long-baseline neutrino experiments with propagation distances of the order of a few hundred kilometers, neutrinos will undergo one or multiple oscillation periods of the DM fields for $m_\phi\gtrsim 10^{-12}\;\mathrm{eV}$. In this regime, the time-dependent perturbation is averaged over a period at the level of the time integral in the evolution operator, so that the contribution that is linear in $\phi$ does not enter the effective Hamiltonian. Thus, in this work, we will not address the DM mass region with $m_\phi\gtrsim 10^{-12}\;\mathrm{eV}$.

\subsection{Effective interactions with neutrinos}
\label{sec:theory2}

Neutrino propagations in the dark medium can subsequently affect the oscillation probabilities. In the case of scalar DM, the relevant Lagrangian can be written as~\cite{Berlin:2016woy}
\begin{equation} \label{eq:lscal}
    {\mathcal L}_{\rm eff}^{\rm scalar} = \bar{\nu}_L^\alpha i \slashed \partial \nu_L^\alpha -\frac{1}{2} m_\nu^{\alpha \beta} (\bar{\nu}_L^c)^\alpha \nu_L^\beta - \frac{1}{2} y^{\alpha \beta} \phi (\bar{\nu}_L^c)^\alpha{\nu_L^\beta},
\end{equation}
where $\nu_L$ is the left-handed neutrino field, $m_\nu$ is the effective neutrino mass matrix, $y$ contains the coupling constants of the neutrino and DM interactions, and $\alpha,\beta$ are the flavor indices for neutrinos.

The scalar field in eq.~(\ref{eq:phi_t}) can have notable implications in neutrino experiments where neutrinos propagate underground. The effective neutrino Lagrangian specified by eq.~(\ref{eq:lscal}) corresponds to the transformation $m_\nu \rightarrow m_\nu + y\phi$, where $m_\nu$ receives an additional contribution $y\phi$ from the neutrino–DM interaction. When the scalar mass $m_\phi$ matches the propagation length of the neutrinos, coherent scattering between the DM and neutrino fields can lead to observable distortions in the oscillation probabilities. Furthermore, the time dependency of $\phi$ means that this distortion varies over time. This modulation may be testable in neutrino experiments where the lifetime is sufficiently long. The oscillation distortion and time modulation of the ULDM scalar fields have previously been contemplated, \textit{e.g.}, in Refs.~\cite{Losada:2021bxx,Losada:2023zap}.

It is also possible to write the neutrino-DM interactions in vector form. In such a case, the DM field alters the neutrino four-momentum as $p_\mu \rightarrow p_\mu + g Q \phi_\mu$, where $\phi_\mu$ represents the vector-like DM field. The corresponding version of the effective Lagrangian is then given by~{\hx \cite{Brdar:2017kbt}}
\begin{equation} \label{eq:lvec}
    {\mathcal L}_{\rm eff}^{\rm vector} = \bar{\nu}_L^\alpha i \slashed \partial \nu_L^\alpha -\frac{1}{2} m_\nu^{\alpha \beta} (\bar{\nu}_L^c)^\alpha \nu_L^\beta + g Q^{\alpha \beta} \phi^\mu \bar{\nu}_L^\alpha \gamma_\mu \nu_L^\beta,
\end{equation}
where $g$ and $Q^{\alpha \beta}$ stand for the effective neutrino-DM coupling and the charge matrix, respectively. The ULDM field $\phi^\mu$ is defined as a vector with its total magnitude given by eq.~(\ref{eq:phi_t}). As noted in Ref.~\cite{Brzeminski:2022rkf}, the vector DM can also be studied through its time-varying effects.

\subsection{Neutrino matter potentials with different scenarios}
\label{sec:theory3}

Neutrino oscillations are affected by the presence of DM particles. The effective Hamiltonian is given by
\begin{equation} \label{eq:5}
    {\mathcal H}_{\rm eff} = \frac{1}{2 E_\nu} \left({U_\nu M^2 U_\nu^\dag}\right) +{V_{\rm CC}} + V_{\rm dark},
\end{equation}
where $M^2 = \diag(0, \Delta m_{21}^2, \Delta m_{31}^2)$, $U_\nu$ is the Pontecorvo-Maki-Nakagawa-Sakata (PMNS) matrix, and $V_{CC}$ denotes the standard matter potential from charged-current interactions. The additional matter potential $V_{\rm dark}$ results from neutrino interactions with DM.

Depending on the Lorentz structures of the neutrino-DM interactions, the DM potential $V_{\rm dark}$ can take different forms. In the case of scalar interactions, this potential can be written as~{\hx \cite{Brdar:2017kbt}}
\begin{equation} \label{eq:6}
        V_{\rm dark}^{\rm scalar} = \frac{1}{2 E_\nu} \left[\phi(y m_\nu + m_\nu y) + \phi^2 y^2 \right],
\end{equation}
where $\phi^2 y^2$ dominates at low values of $m_\phi$, while $\phi(y m_\nu + m_\nu y)$ at high values of $m_\phi$. It should be noted that the DM affects neutrino oscillations in two different ways: the time modulation and the non-trivial matter effects. The time modulation arises from the classical field definition (c.f. eq.~\ref{eq:phi_t}) and the strength of the modulation depends primarily on the neutrino time of flight. The matter effect, on the other hand, is influenced by the matter potential $V_{\rm dark}^{\rm scalar}$, which can be calculated from eq.~(\ref{eq:6}).

The couplings between neutrinos and the scalar ULDM are given by the $3\times3$ matrix $y$, which is defined as
\begin{equation} \label{eq:7}
    y = \left(
            \begin{array}{ccc}
                y_{ee} & y_{e\mu} & y_{e\tau} \\[0.1cm]
                y_{ee}^* & y_{\mu\mu} & y_{\mu\tau} \\[0.1cm]
                y_{e\tau}^* & y_{\mu\tau}^* & y_{\tau\tau}
            \end{array}
        \right),
\end{equation}
where $y_{e e}$, $y_{\mu \mu}$ and $y_{\tau \tau}$ are real, while $y_{e \mu}$, $y_{e \tau}$ and $y_{\mu \tau}$ are complex parameters. These couplings can follow a specific pattern, as is the flavor-universal case with $y_{ee} = y_{\mu\mu} = y_{\tau\tau}$. The coupling parameters may also be unstructured, in which case any of the diagonal and non-diagonal parameters can acquire different non-zero values. We define the off-diagonal parameters as $y_{\ell \ell'} = |y_{\ell \ell'}| e^{-i \phi_{\ell \ell'}}$, where $|y_{\ell \ell'}|$ and $\phi_{\ell \ell'}$ represent the magnitude and complex phase of the given coupling, for which $\ell$, $\ell' = e$, $\mu$ or $\tau$.

In DM models that involve vector-like interactions, the neutrino matter potential is expressed in terms of the charge matrix $Q$ and the effective neutrino-DM coupling $g$ as~{\hx \cite{Brdar:2017kbt}}
\begin{equation} \label{eq:8}
    V_{\rm dark}^{\rm vector} = -\frac{1}{2E_\nu} \left[2 (p_\nu \cdot \phi) g Q + g^2 Q^2 \phi^2 \right].
\end{equation}
Although its elements can in principle take any values, the charge matrix $Q$ often has, in practice, a structure that follows from the DM model. In this work, we are mainly interested in the DM candidates that are charged under the $L_e - L_\mu$ or $L_\mu - L_\tau$ gauge symmetry. In the former case, the charge matrix is defined as $Q = \diag(1, -1, 0)$, while in the latter case $Q = \diag(0, 1, -1)$. For clarity, we denote the effective coupling by $g$ in the $L_e - L_\mu$ case and by $g'$ in the $L_\mu - L_\tau$ case, respectively.

In this work, we consider the scalar ULDM in the mass range $m_\phi \in [10^{-23}, 10^{-12}]$\,eV and study its effects on long-baseline neutrino oscillations with both flavor-universal and flavor-general couplings. We also investigate the effects of the vector ULDM fields in the $L_e - L_\mu$ and $L_\mu - L_\tau$ cases, with a similar mass range.

\section{Analysis of the long-baseline neutrino oscillation data}
\label{sec:data}

In this work, we investigate two different datasets from the on-going long-baseline neutrino oscillation experiments. The first dataset comes from the T2K experiment~\cite{T2K:2023smv}, which is currently taking data in Japan. The second dataset is taken from NO$\nu$A~\cite{NOvA:2025tmb} in USA. In this section, we briefly introduce the experimental setups of T2K and NO$\nu$A and describe the analysis methods used in this work.

\subsection{T2K dataset}
\label{sec:data:T2K}

The T2K is a long-baseline accelerator neutrino experiment currently taking data in Japan. It operates a proton linac with $770$~kW average power output. The accelerator produces a high-purity neutrino beam, of which $96\%$-$98\%$ are muon neutrinos or muon antineutrinos. The experiment also has three detector facilities. The main detector, Super-Kamiokande, is a large water Cherenkov vessel located $295$~km from the accelerator with a $2.5^\circ$ off-axis angle. The other two detectors, ND280 and INGRID, are placed near the accelerator to monitor the neutrino beam and constrain its beam-related systematic uncertainties. The T2K produces a narrow-band beam, which peaks at about $600$~MeV in the far detector facility. The main goal of T2K experiment is to find evidence for leptonic CP violation and improve measurements of other neutrino oscillation parameters and neutrino-nucleus cross-sections~\cite{T2K:2023smv}. 

We analyze the latest neutrino oscillation dataset from the T2K experiment~\cite{T2K:2023smv}, which accounts for neutrino events from $1.97\times10^{21}$ protons-on-target (POT) in neutrino mode and $1.63\times10^{21}$ POT in antineutrino mode, respectively. The dataset comprises five interaction samples: the charged-current quasi-elastic (CCQE) $\nu_\mu$ and $\nu_e$ events obtained in neutrino mode, the CCQE $\bar{\nu}_\mu$ and $\bar{\nu}_e$ events in antineutrino mode, and the resonant charged-current pion production (CC1$\pi$) $\nu_e$ events in neutrino mode. The neutrino and antineutrino events that make up these samples correspond to the oscillation channels $\nu_\mu \rightarrow \nu_e$, $\nu_\mu \rightarrow \nu_\mu$, $\bar{\nu}_\mu \rightarrow \bar{\nu}_e$, and $\bar{\nu}_\mu \rightarrow \bar{\nu}_\mu$. We analyze neutrino and antineutrino events assuming $36$ energy bins over the $[0.2, 2.0]$~GeV interval in both $\nu_\mu$ and $\bar{\nu}_\mu$ CCQE samples, and $23$ bins within the $[0.1, 1.25]$~GeV energy range for $\nu_e$ and $\bar{\nu}_e$ CCQE samples, respectively. We also include the $\nu_e$CC1$\pi^+$ sample events with $19$ energy bins over the $[0.45, 1.25]$~GeV energy range.

\subsection{\texorpdfstring{NO$\nu$A}{NOnuA} dataset}
\label{sec:data:NOvA}

The NO$\nu$A is a long-baseline neutrino oscillation experiment currently operating in USA. Akin to the T2K experiment, the NO$\nu$A generates intensive beams of muon neutrinos and antineutrinos using accelerated protons. The neutrino beam facility NuMI operates at $700$~kW average beam power and its output is monitored by a near detector located $1$~km away from the beamline. The far detector is located $810$~km from the source and is a segmented active calorimeter consisting of highly reflective PVC cells. The total fiducial mass of the far detector is $14$~kt. Both detector facilities are placed $0.8^\circ$ off-axis. The NO$\nu$A beam facility generates a wide range of energies with $2$~GeV average. The main objectives of NO$\nu$A experiment include probing the violation or conservation of leptonic CP and establishing the correct ordering of neutrino masses~\cite{NOvA:2025tmb}.

The NO$\nu$A dataset~\cite{NOvA:2025tmb} consists of six event samples that correspond to $26.6\times$10$^{20}$ POT in neutrino mode and $12.5\times$10$^{20}$ POT in antineutrino mode, respectively. Due to the wide neutrino energy range, neutrino events belonging to the NO$\nu$A dataset include various types of neutrino charged-current interactions. The first four samples are the electron-like events that undergo $\nu_\mu \rightarrow \nu_e$ and $\bar{\nu}_\mu \rightarrow \bar{\nu}_e$ oscillations. Depending on the purity level of particle identification used in the convolutional neural network (CNN) algorithm, the electron-like samples are further divided into the so-called low-CNN$_{\rm evt}$ and high-CNN$_{\rm evt}$ bins. The remaining two data samples correspond to the muon-like events that characterize the disappearance channels $\nu_\mu \rightarrow \nu_\mu$ and $\bar{\nu}_\mu \rightarrow \bar{\nu}_\mu$. Whereas the electron-like events are distributed in $6$ same-sized bins over $[1.0, 4.0]$~GeV energies, the muon-like events are binned over $[0.0, 5.0]$~GeV energy range with $22$ bins of unequal size.

\subsection{Data analysis}
\label{sec:data:methods}

The analysis of the T2K and NO$\nu$A datasets is performed with a modified version of the General Long-Baseline Experiment Simulator (GLoBES)~\cite{Huber:2004ka,Huber:2007ji}. To this end, we have altered the GLoBES package to calculate oscillation probabilities with scalar and vector ULDM parameters discussed in section~\ref{sec:theory2}. The full description of the analysis methods can be found in Ref.~\cite{Lin:2021cst}.

The GLoBES package calculates the neutrino events by integrating fluxes, cross-sections, and oscillation probabilities over neutrino energies{\hx ~\cite{Huber:2004ka}}:
\begin{equation} \label{eq:LBLevent}
    N_i = N_{\rm nucl} \, T \, \epsilon \int_{E_{\rm min}}^{E_{\rm max}}  \int_{E_{\rm min}^{'}}^{E_{\rm max}^{'}} dE \, dE^{'} \, {\hx \Phi_\nu(E)} \, \sigma(E) \, R(E, E^{'}) \, P_{\nu_{\ell} \rightarrow \nu_{\ell'}}(E),
\end{equation}
where the indices $i = 1, 2, 3, \ldots$ run over the neutrino energy bins, and ${\hx \Phi_\nu(E)}$, $\sigma(E)$, and $R(E, E^{'})$ represent the neutrino fluxes, the neutrino-nucleus cross-sections, and the energy resolution function, respectively. The oscillation probabilities $P_{\nu_{\ell} \rightarrow \nu_{\ell'}}(E)$ are computed as a function of the true neutrino energy $E$, including both matter effects and neutrino-DM interactions. The double integration in eq.~(\ref{eq:LBLevent}) is carried out over the true neutrino energy $E$ and the reconstructed neutrino energy $E'$. The final result is multiplied by $N_{\rm nucl}$, $T$, and $\epsilon$, which represent the number of nucleons in the detector, the total lifetime of the experiment, and the efficiency of the detector, respectively.

The analysis of the T2K and NO$\nu$A datasets is conducted with the following formula~{\cite{Huber:2007ji,Fogli:2002pt}}:
\begin{equation} \label{LBLChi2}
\chi^2  = \sum_{i} 2\left[ T_{i} - O_{i} \left( 1 + \ln\frac{O_{i}}{T_{i}} \right) \right]
 + \frac{\zeta_{\text{sg}}^2}{\sigma_{\zeta_{\text{sg}}}^2} + \frac{\zeta_{\text{bg}}^2}{\sigma_{\zeta_{\text{bg}}}^2} + {\rm priors},
\end{equation}
where $T_{i}$ and $O_i$ represent the theoretically predicted and experimentally observed events, respectively. $\zeta_{\text{sg}}$ and $\zeta_{\text{bg}}$ are the nuisance parameters representing systematic errors in signal and background normalizations, with the corresponding $1\sigma$ confidence-level (CL) uncertainties given by $\sigma_{\zeta_{\text{sg}}}$ and $\sigma_{\zeta_{\text{bg}}}$, respectively. The predicted events $T_i$ are obtained from eq.~(\ref{eq:LBLevent}) as $T_i = (1 + \zeta_{\text{sg}}) N_i^{\text{sg}} + (1 + \zeta_{\text{bg}}) N_i^{\text{bg}}$, where $N_i^{\text{sg}}$ and $N_i^{\text{bg}}$ denote the signal and background events, respectively. We adopt $5\%$ signal and $10\%$ background errors for the T2K dataset, and $5\%$ errors for signal and background events in the NO$\nu$A dataset. {\hx These choices of systematic uncertainties are consistent with those in Ref.~\cite{Lin:2021cst}. We have also confirmed that they can reproduce the numerical results reported in Refs.~\cite{T2K:2023smv,NOvA:2025tmb}.}

The stochastic fluctuations are incorporated for the ULDM within a Bayesian framework~\cite{Derevianko:2016vpm,Gregory:2005zz}. The marginalized likelihood chi-square with $N_{\rm DM}=\lfloor T_{\rm exp}/\tau_c \rfloor + 1$ independent samples can be written as
\begin{equation} \label{eq:chi2_stoch}
    \chi_{\rm stoch}^2(y) = -2\ln \int 
    \exp\!\left[-\frac{1}{2}\chi^2(y,\phi_0^1,\dots,\phi_0^{N_{\rm DM}})\right]
    \prod_{i=1}^{N_{\rm DM}} p(\phi_0^i)\, d\phi_0^i .
\end{equation}
Since the stochastic field amplitude affects observables only via the energy density $\rho_\phi \propto \phi^2$, which enters through the product $y\sqrt{\rho_\phi}$ with the coupling constant $y$, the expression in eq.~(\ref{eq:chi2_stoch}) can be simplified by introducing a dimensionless and normalized density variable $x = \overline{\rho_\phi}/\rho_{\rm DM}$ as
\begin{equation}\label{eq:chi2_stoch_simp}
    \chi_{\rm stoch}^2(y) = -2\ln \int_0^\infty
    \exp\!\left[-\frac{1}{2}\chi^2_{\rm det}(y\sqrt{{x}},\rho_{\rm DM})\right]
    \Gamma(x; N_{\rm DM},\, 1/N_{\rm DM})\;dx,
\end{equation}
where $\Gamma(x;k,\theta)$ denotes a Gamma distribution with shape parameter $k$ and scale parameter $\theta$. Here $\chi^2_{\rm det}(y\sqrt{{x}},\rho_{\rm DM})$ denotes the deterministic chi-square computed for a fixed (non-stochastic) field amplitude, evaluated at an effective coupling $y\sqrt{x}$. The simplification to eq.~(\ref{eq:chi2_stoch_simp}) is possible because, under the classical field approximation, the coupling constant always appears in the Hamiltonian multiplied by the DM field amplitude (\textit{i.e.}, $y\phi_0$ in the scalar case and $g\phi_0$ in the vector case). Since $\phi_0\propto\sqrt{\rho_\phi}$, eqs.~(\ref{eq:chi2_stoch}) and (\ref{eq:chi2_stoch_simp}) take a similar form for the vector DM, with the simple replacement $y \rightarrow g$.

In the limit where the experimental exposure time is much longer than the coherence time, $N_{\rm DM}\gg 1$, $x$ will concentrate around its expectation value $x = 1$. The central limit theorem then implies that the width of the distribution shrinks as $N_{\rm DM}^{-1/2}$, so that the Gamma distribution becomes sharply peaked and approaches a Dirac delta function,
\begin{equation}
     \Gamma(x;N_{\rm DM},1/N_{\rm DM}) \;\longrightarrow\; \delta(x-1).
\end{equation}
Consequently, eq.~(\ref{eq:chi2_stoch}) reduces exactly to the deterministic likelihood $\chi^2_{\rm det}(y,\rho_{\rm DM})$.

The combined neutrino oscillation datasets from T2K and NO$\nu$A are analyzed with the following restrictions on the standard oscillation parameters. The solar parameters $\theta_{12}$ and $\Delta m_{21}^2$ are fixed at 33.76$^\circ$ and 7.54$\times$10$^{-5}$eV$^2$ respectively, adopting the best-fit values from the NuFit-6.1 database~\cite{Esteban:2020cvm,NuFIT:6-1}, as their uncertainties have a negligible impact on the observables in T2K and NO$\nu$A. Furthermore, we apply a prior to the reactor mixing angle $\theta_{13}$ by demanding that $\sin^2 2\theta_{13} =$ 0.0851$\pm$0.0024 at 1$\sigma$ CL~\cite{DayaBay:2022orm}. The $\chi^2$ function defined in eq.~(\ref{LBLChi2}) is minimized over the atmospheric parameters $\theta_{23}$ and $\Delta m_{31}^2$ and also the Dirac CP-violating phase $\delta_{CP}$. Throughout this work, we assume normally ordered neutrino masses, which are currently favored by the experimental data~\cite{NuFIT:6-1}.

\section{Numerical results}
\label{sec:results}

We now present our results from a numerical analysis of the combined T2K and NO$\nu$A datasets, first for the scalar DM in section~\ref{sec:results:scalar} and then for the vector DM in the $L_e - L_\mu$ and $L_\mu - L_\tau$ cases in section~\ref{sec:results:vector}.

\subsection{Ultralight scalar dark matter}
\label{sec:results:scalar}

\begin{figure}[t]
    \centering
    \includegraphics[width=0.65\textwidth]{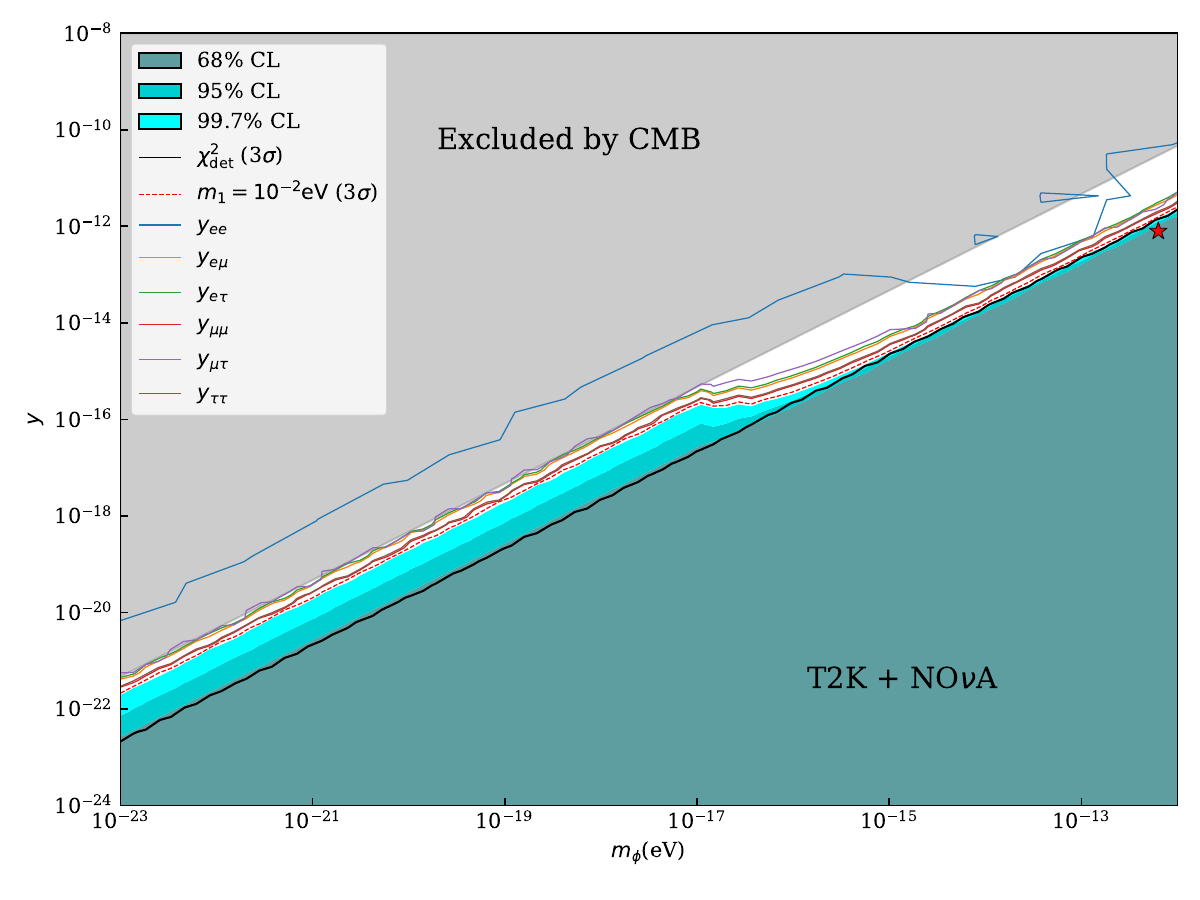}
    \caption{Joint T2K and NO$\nu$A constraints on the scalar DM. The color map shows the $\chi^2$ distribution for the flavor-universal scenario, with the black solid and red dashed lines representing the $3\sigma$ boundaries for the deterministic and the $m_0 = 10^{-2}$~eV case, respectively. The individual $3\sigma$ upper limits for general scalar couplings are shown as solid curves ($y_{\alpha\beta}$). The grey region indicates the conservative CMB exclusion~\cite{Krnjaic:2017zlz}. The best-fit point is located at $m_{\phi} \simeq 6.31 \times 10^{-13}~\text{eV}$ and $y_0 \simeq 7.93 \times 10^{-13}$. Unless noted otherwise, the normal ordering and $m_1 = 0$ are assumed for neutrinos.\label{fig:scalar_DM}}
\end{figure}

We analyze the joint datasets from the T2K and NO$\nu$A experiments within the framework of scalar DM. Two distinct coupling scenarios are explored: a flavor-universal pattern and a set of general (non-universal) patterns. The resulting constraints on the coupling strengths $y_{\alpha\beta}$ and the DM mass $m_{\phi}$, spanning the range $10^{-23}~\text{eV} \leq m_{\phi} \leq 10^{-10}~\text{eV}$, are summarized in Figure~\ref{fig:scalar_DM}.  

In the flavor-universal scenario, the neutrino-DM couplings are assumed to be flavor-blind, such that the coupling matrix takes the form $y = y_0\,\mathbb{I}_{3\times3}$, where $y_0$ denotes the universal coupling strength and $\mathbb{I}_{3\times3}$ the $3\times3$ identity matrix. This benchmark case is well-motivated by flavor symmetries and has been extensively investigated in the context of neutrino oscillation experiments~\cite{Berlin:2016woy,Capozzi:2017auw,Brdar:2017kbt}. Conversely, for the general coupling patterns, we adopt a parameter-by-parameter approach, varying a single coupling constant $y_{\alpha\beta}$ at a time, while setting all other off-diagonal and diagonal entries to zero. Since $y_{ee}$ effectively contributes as an additional term to the matter potential and is therefore weakly constrained by the T2K and NO$\nu$A datasets, the resulting exclusion curves exhibit some spurious features. The leading experimental constraint in this ULDM mass regime arises from the CMB measurements~\cite{Krnjaic:2017zlz}, which set a stringent upper limit on the sum of neutrino masses $\sum m_{\nu}$. To be consistent with the analysis performed in Ref.~\cite{Brdar:2017kbt}, we reproduce this cosmological bound by adopting the conservative criteria detailed in Ref.~\cite{Krnjaic:2017zlz}.

As illustrated in eq.~\ref{eq:6}, the effective potential induced by the scalar DM field contains a cross-term proportional to the neutrino mass, rendering the interaction sensitive to the absolute neutrino mass scale. Given the well-measured mass-squared differences, the entire neutrino mass spectrum can be uniquely determined by specifying the mass of the lightest neutrino ($m_1$ in the case of normal ordering). Our benchmark analysis assumes a vanishing lightest neutrino mass ($m_1 = 0$), a configuration that is theoretically well-motivated by the minimal see-saw model~\cite{King:1999mb,Frampton:2002yf,King:2013iva}.

To evaluate sensitivity to the absolute mass scale, we also investigate a scenario with $m_1 = 1.0 \times 10^{-2}$~eV (represented by the red dashed line in Fig.~\ref{fig:scalar_DM}), a value that approaches the current experimental upper bounds. Our results indicate that the specific choice of the lightest neutrino mass does not significantly alter the overall constraints, resulting only in a marginal relaxation of the excluded regions.

In our statistical framework, the stochastic nature of the scalar DM field is rigorously incorporated. Given a characteristic experimental exposure time ($T_{\rm exp}$) of approximately ten years, the stochastic fluctuations manifest most significantly in the ULDM mass regime of $m_{\phi} \lesssim 10^{-17}$~eV, where the coherence time $\tau_c$ is comparable to or exceeds $T_{\rm exp}$. In this maximum fluctuation region, the $3\sigma$ exclusion limit for the flavor-universal scenario approximately follows the scaling relation $y_0 / m_{\phi} \simeq 19.4~\text{eV}^{-1}$.

As the DM mass $m_{\phi}$ increases, the coherence time $\tau_c \propto (m_{\phi} v^2)^{-1}$ decreases, leading to a higher number of independent stochastic samples ($N_{\rm DM} \simeq T_{\rm exp}/\tau_c$) within the observation window. This results in a statistical averaging of the field's fluctuations, making the constraints progressively converge toward the deterministic limit. In this high-mass regime where fluctuations are fully averaged, the $3\sigma$ boundary follows a more stringent scaling of $y_0 / m_{\phi} \simeq 2.26~\text{eV}^{-1}$. The discrepancy between these two regimes quantifies the order-of-magnitude relaxation (a factor of $\sim 8.6$) in the sensitivity induced by the stochastic nature of the net DM field.

Finally, a global scan across both flavor-universal and general coupling patterns yields a best-fit point at $m_{\phi} \simeq 6.31 \times 10^{-13}~\text{eV}$ and $y_0 \simeq 7.93 \times 10^{-13}$. However, this result exhibits less than $1\sigma$ improvement ($\Delta\chi^2\sim0.4$) relative to the standard neutrino oscillation hypothesis (\textit{i.e.}, excluding the DM background), indicating that there is no statistically significant preference for the scalar DM model by the current T2K and NO$\nu$A datasets.

\begin{figure}[t]
    \centering
    \includegraphics[width=0.60\linewidth]{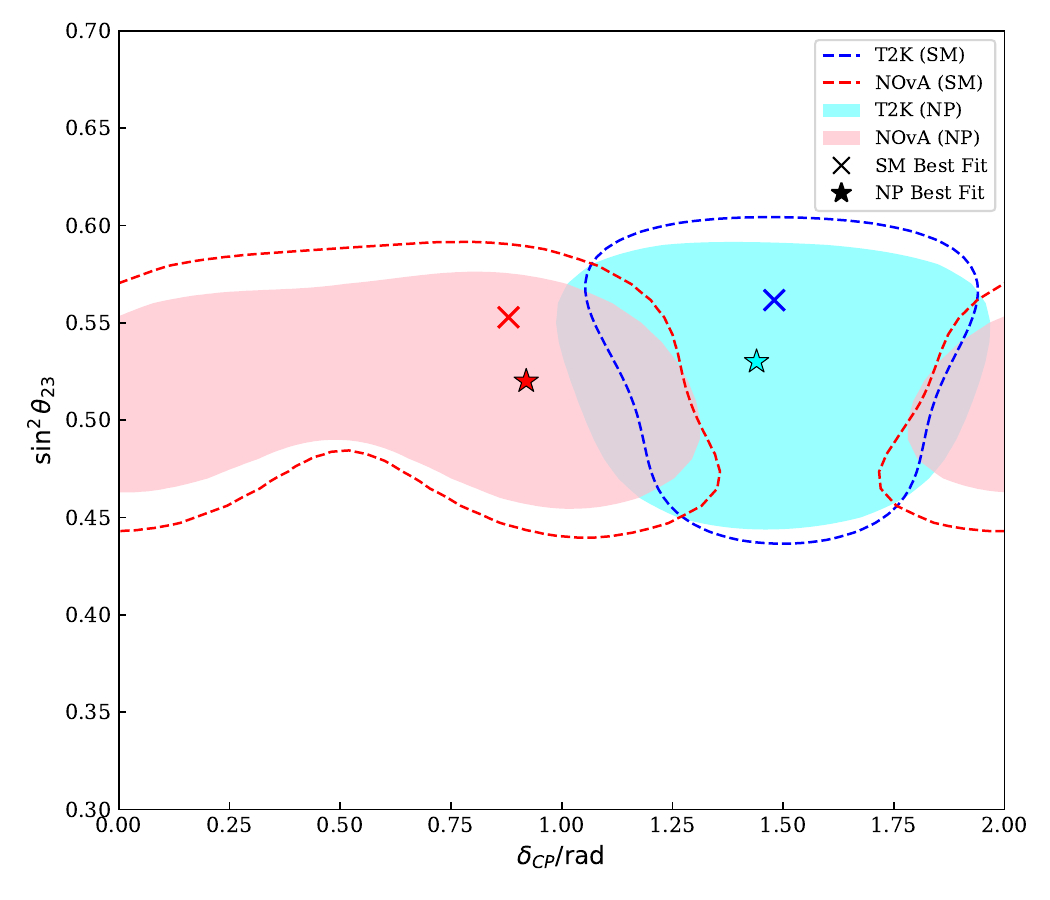}
    \caption{Effects of the flavor-universal scalar ULDM in the combined analysis of T2K and NO$\nu$A datasets. The results are presented at $90\%$ CL for the normal neutrino mass ordering with the assumption of $m_1 = 0$. The new physics (NP) contours (shaded regions) are computed at the scalar DM best-fit point, corresponding to $m_{\phi} \simeq 6.31 \times 10^{-13}~\text{eV}$ and $y_0 \simeq 7.93 \times 10^{-13}$. Normal ordering and $m_1 = 0$ are assumed for neutrinos.}
    \label{fig:scalar_exp}
\end{figure}

Next, we investigate the effects of neutrino-DM couplings in the measurement of standard parameters $\sin^2 \theta_{23}$ and $\delta_{CP}$. Figure~\ref{fig:scalar_exp} illustrates the joint fit results for $\sin^2 \theta_{23}$ and $\delta_{CP}$ at $90\%$ CL, with the analysis presented separately for T2K (blue) and NO$\nu$A (red). The dashed contours represent the standard oscillation framework, while the shaded regions denote the flavor-universal scalar DM scenario. Incorporating the DM coupling leads to a slight downward shift in the preferred values of $\sin^2 \theta_{23}$ for both T2K and NO$\nu$A. Additionally, the best-fit regions for $\delta_{CP}$ show marginal convergence to each other compared to the SM predictions. However, the overall parameter spaces for both experiments remain largely consistent with their SM counterparts. Consequently, based on the current available data, the scalar ULDM model does not appear to provide a significant contribution toward alleviating the existing tension between T2K and NO$\nu$A.

\subsection{Ultralight vector dark matter with \texorpdfstring{$L_e - L_\mu$}{LemLmu} and \texorpdfstring{$L_\mu - L_\tau$}{LmumLtau}}
\label{sec:results:vector}

The ULDM may also couple with active neutrinos via vector interactions. In such a case, the effects on neutrino oscillations are induced by the DM polarization, which is not present in the scalar case. Thus, the vector ULDM may be studied in neutrino oscillation experiments by daily modulation~\cite{Brzeminski:2022rkf}. As Earth's rotation changes the angle at which neutrino and vector DM may interact, the subsequent effect on the neutrino matter potential also depends on the time at which the interaction occurs. {\hx In this work, we focus primarily on the polarized vector DM scenario, while also extending our analysis to the unpolarized scenario. Physically, the unpolarized scenario characterizes a regime dominated by multi-mode random wave interference, producing a robust isotropic background due to high particle occupations. In the polarized scenario, on the other hand, it is assumed that the vector DM field possesses a specific preferred orientation, for which we conservatively adopt the configuration that maximizes the interaction strength as a phenomenological benchmark.}

Constraints on the interactions of vector ULDM with neutrinos have been derived previously, with the leading constraints coming from black hole superradiance~\cite{Baryakhtar:2017ngi} and atmospheric neutrinos~\cite{Brzeminski:2022rkf}. Other sources of experimental constraints include the search for the fifth force~\cite{Wise:2018rnb,Coloma:2020gfv}, the cosmological bounds on neutrino decay~\cite{Chen:2022idm}, and the tests of the equivalence of gravitational mass and inertial mass~\cite{Schlamminger:2007ht}. There is also a model-dependent constraint from searches for mono-leptons and missing energy at the Large Hadron Collider (LHC)~\cite{Ekhterachian:2021rkx}. These constraints have been collected in Ref.~\cite{Brzeminski:2022rkf} and are reproduced in the present work.

\begin{figure}[t]
    \centering
    \includegraphics[width=0.65\textwidth]{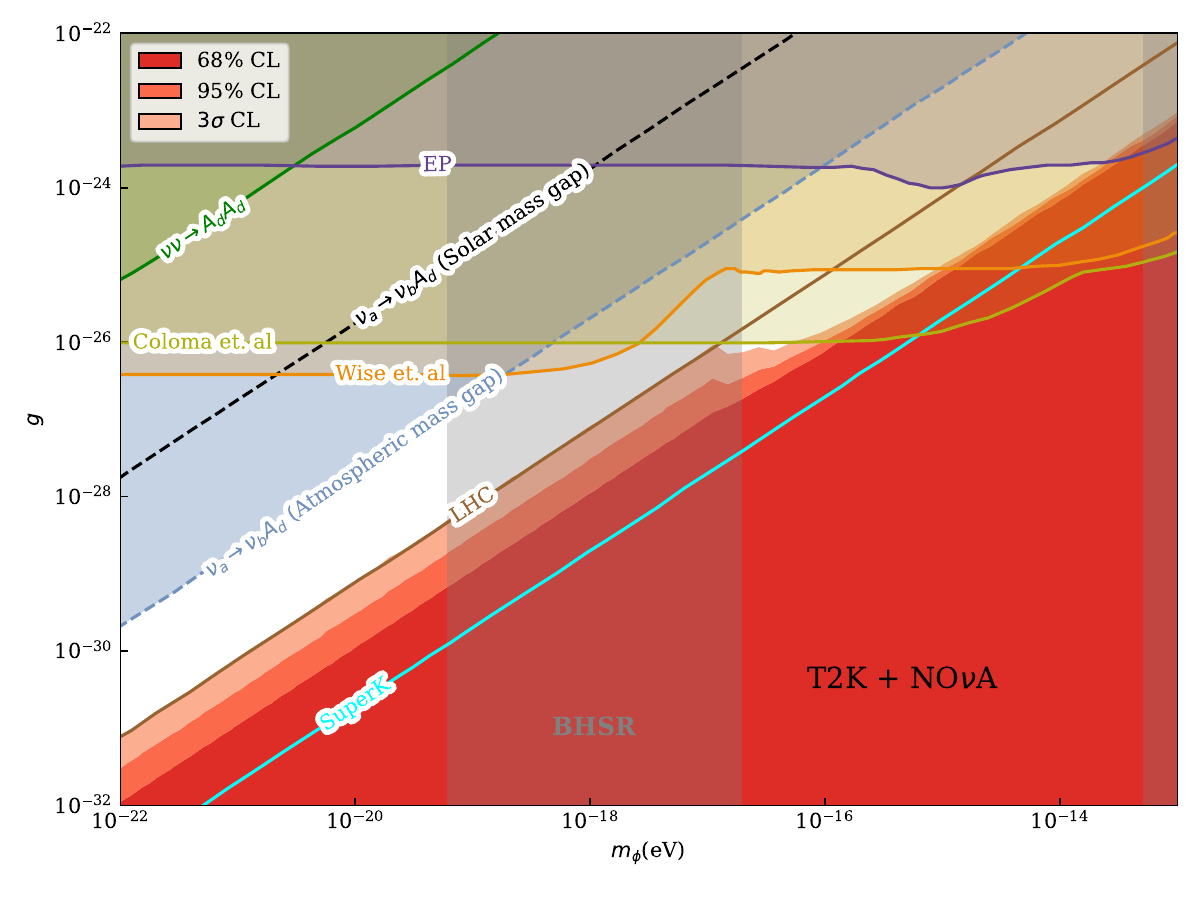}
    \caption{Vector DM fit with neutrino-DM coupling $g$ under the $L_e - L_\mu$ charge. The allowed values for $g$ and $m_\phi$ arising from the joint T2K and NO$\nu$A datasets are presented at $68\%$, $95\%$ and $3\sigma$ CL. External constraints are taken from Ref.~\cite{Brzeminski:2022rkf}. The best-fit point, located at $g \simeq 7.93\times10^{-24}$ and $m_\phi \simeq 1.58\times 10^{-13}$ eV, lies near the right boundary of the parameter space shown, assuming the normal mass ordering. \label{fig:vector_Le-Lmu_DM}}
\end{figure}

We begin by analyzing the vector DM scenario, where the model is charged under the $L_e - L_\mu$ gauge symmetry. Figure~\ref{fig:vector_Le-Lmu_DM} shows the fit results under the combined constraints from T2K and NO$\nu$A datasets in this scenario. The coupling parameter $g$ represents the coupling strength between neutrinos and the vector DM, whereas the charge matrix is given as $Q = \diag(+1, -1, 0)$. The allowed values of $g$ and $m_\phi$ are shown for $68\%$, $95\%$ and $3\sigma$ CL in the T2K and NO$\nu$A datasets with the red, dark red and light red regions, respectively. 

The fit is presented for the polarized DM case, where its effect on the neutrino matter potential is maximal. The figure also shows the present experimental constraints from the atmospheric neutrinos (SuperK), the fifth force searches~\cite{Wise:2018rnb,Coloma:2020gfv}, the cosmological limits from solar and atmospheric masses ($\nu_a \rightarrow \nu_b A_d$ for the solar and $\nu_a \rightarrow \nu_b A_d$ for the atmospheric mass gap, respectively), and the equivalence principle (EP) tests~\footnote{Violation of the equivalence principle has also been studied in MICROSCOPE space mission, which sets an independent upper limit of $1.3\times10^{-25}$ on the couplings relevant for the $L_e - L_\mu$ gauge symmetry~\cite{Fayet:2017pdp,Fayet:2018cjy}.}. We also show the limits from the LHC and black hole superradiance (BHSR).

In terms of experimental sensitivity, the combined constraints from T2K and NO$\nu$A on the $L_e - L_\mu$ parameter space fall between the limits set by the LHC and the Super-Kamiokande atmospheric neutrino data. The best-fit point exhibits a marginal improvement in the fit ($\Delta\chi^2\sim0.5$ within the $1\sigma$ CL) and is situated in a region already excluded by existing experimental bounds. Notably, our T2K and NO$\nu$A limits appear relatively relaxed in the low-mass regime due to the incorporation of stochastic fluctuations. It is important to emphasize that if other external constraints were consistently treated under the classical field framework, they would similarly be subject to such stochastic effects, leading to a corresponding reassessment of their exclusion boundaries.

We further explore the $L_e - L_\mu$ case within the unpolarized scenario. In this regime, the first-order correction to the neutrino matter potential is averaged out, making only the second-order terms provide a nonvanishing contribution; consequently, the resulting effect on neutrino oscillation probabilities is significantly suppressed compared to the polarized case. Due to this weakened sensitivity, the constraints on unpolarized DM are considerably relaxed. The corresponding best-fit point fails to yield a significant improvement in the fit and is subsequently excluded by all experimental constraints considered in this work.

\begin{figure}[t]
    \centering
    \includegraphics[width=0.65\textwidth]{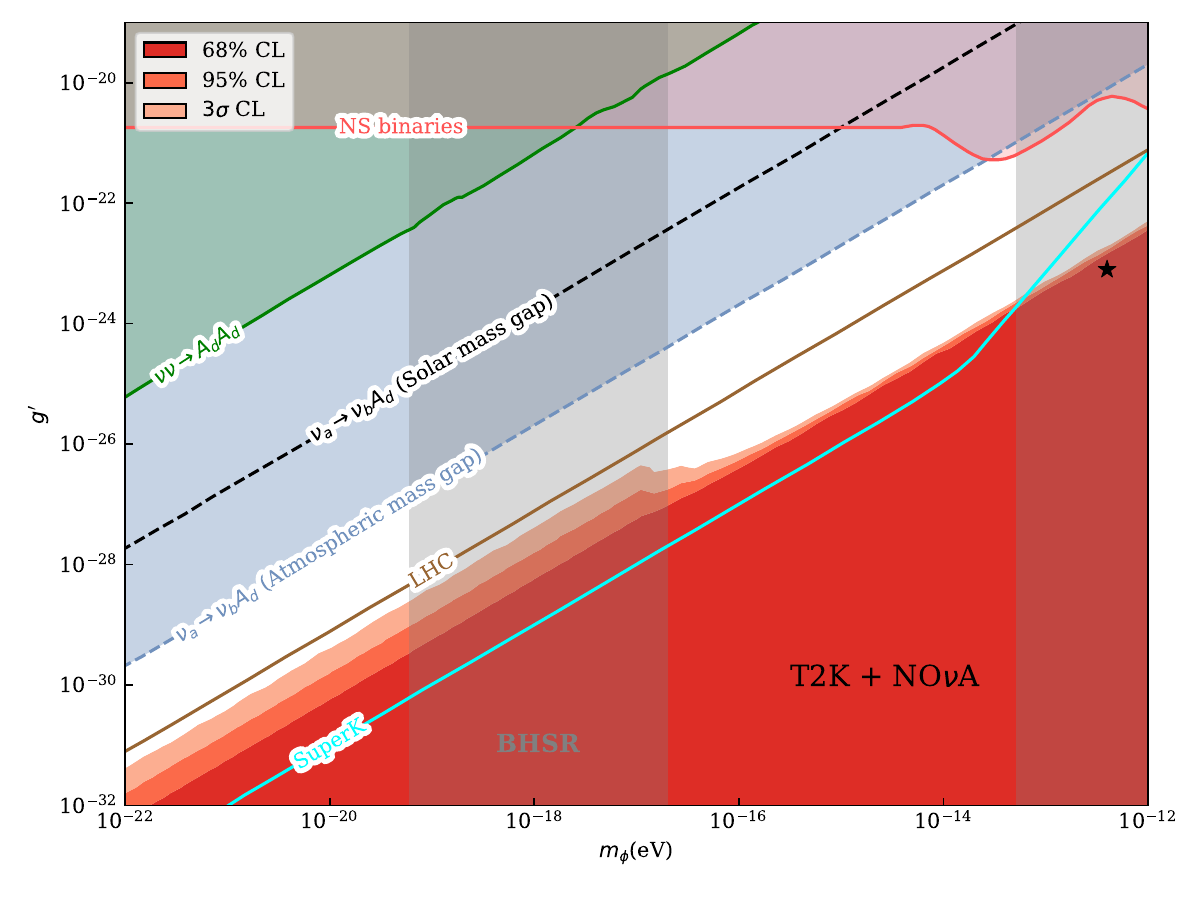}
    \caption{Vector DM fit with neutrino-DM coupling $g'$ under the $L_\mu - L_\tau$ charge. The allowed values of $g'$ and $m_\phi$ are presented for the combined T2K and NO$\nu$A datasets at $68\%$, $95\%$ and $3\sigma$ CL, assuming the normal mass ordering. Constraints from other experiments are taken from Ref.~\cite{Brzeminski:2022rkf}. \label{fig:vector_Lmu-Ltau_DM}}
\end{figure}

Next, we examine the combined T2K and NO$\nu$A datasets in the $L_\mu - L_\tau$ case. The results are shown in Figure~\ref{fig:vector_Lmu-Ltau_DM}. In this case, the charge matrix of the ULDM field $\phi$ is given by $Q = \diag(0, +1, -1)$. The leading experimental constraints are the same as those in the $L_e - L_\mu$ case, except for the bounds coming from the fifth force and the violation of EP searches, which are replaced with constraints from the neutron star (NS) binaries~\cite{Dror:2019uea}. The allowed values for the T2K and NO$\nu$A datasets are presented at $68\%$, $95\%$ and $3\sigma$ CL by red, dark red and light red regions, and the best-fit point by black star. The fit results are obtained for the polarized DM.

At $3\sigma$ CL, the neutrino oscillation datasets from T2K and NO$\nu$A experiments provide stringent constraints on the vector ULDM parameter space in the $L_\mu - L_\tau$ sector. Specifically, in the low-mass regime where stochastic fluctuations are maximal, we find $g'/m_\phi \lesssim 4.08 \times 10^{-10}~\text{eV}^{-1}$. In the high-mass region where fluctuations are effectively averaged out, on the other hand, the constraint is tightened to $g'/m_\phi \lesssim 4.82 \times 10^{-11}~\text{eV}^{-1}$. The best-fit point, denoted by a black star in the figure, is located at $m_\phi \simeq 3.98 \times 10^{-13}~\text{eV}$ and $g' \simeq 7.93 \times 10^{-24}$. These limits are slightly less stringent than those reported in Ref.~\cite{Alonso-Alvarez:2024wnh}, although they remain consistent in terms of orders of magnitude. This discrepancy primarily stems from our explicit treatment of stochastic fluctuation effects and the incorporation of phase averaging within the oscillation probabilities, as discussed in section~\ref{sec:theory1}.

Similarly to the $L_e - L_\mu$ case, the BHSR bounds truncate certain portions of the parameter space while leaving several regions of interest intact, and the best-fit point fails to yield a statistically significant improvement ($\Delta\chi^2\sim0.2$) to the overall fit. The results for the unpolarized scenario are omitted from the figure as they are robustly excluded and their corresponding limits are entirely overshadowed by other constraints.

\begin{figure}[t]
    \centering
    \includegraphics[width=\textwidth]{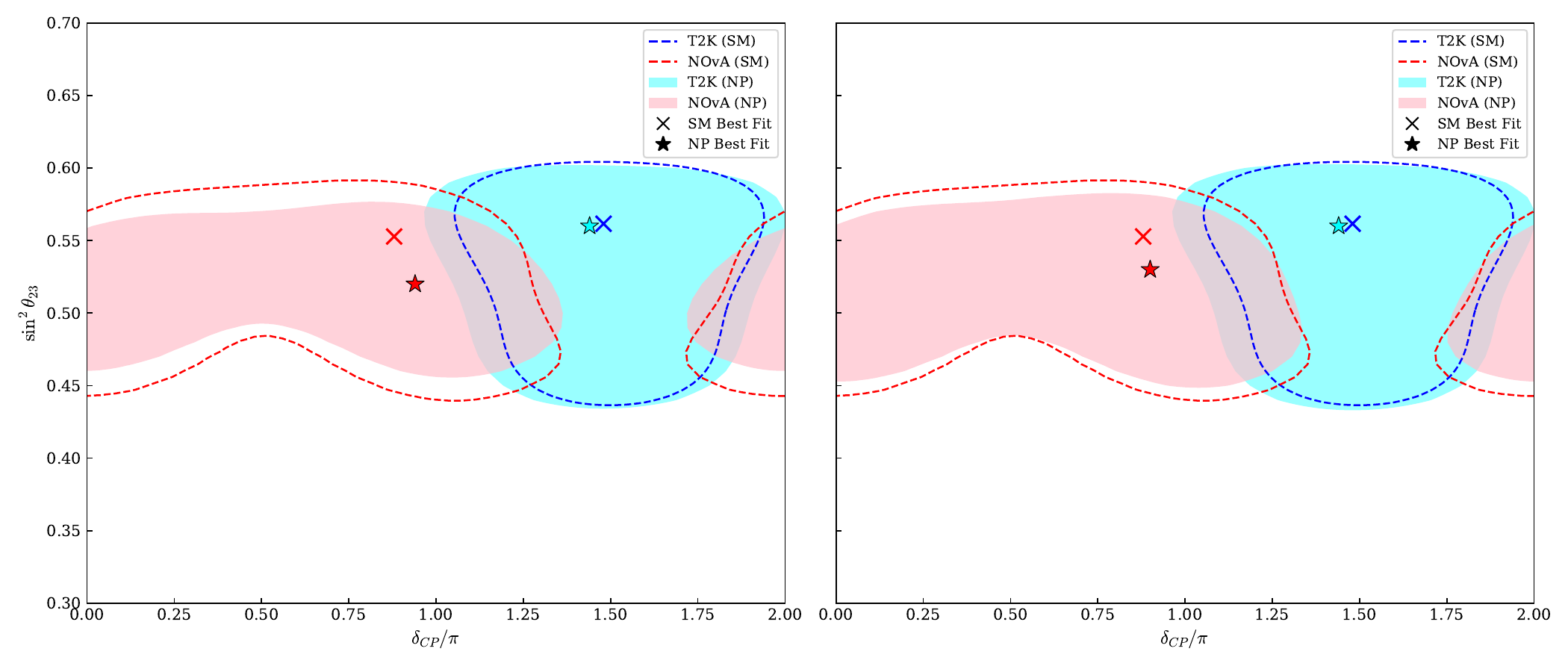}
    \caption{Discrepancies in T2K and NO$\nu$A fits for the vector ULDM with $L_e - L_\mu$ (left panel) and $L_\mu - L_\tau$ (right panel) charges. The vector couplings $g$ and $g'$ and DM mass $m_\phi$ are fixed to their best-fit points in both panels. Sensitivities are presented for T2K and NO$\nu$A at $90\%$ CL assuming normally ordered neutrino masses. \label{fig:vector_exp}}
\end{figure}

As the last step of this work, we study the effect of the vector ULDM on the parameter discrepancy in the T2K and NO$\nu$A datasets. Figure~\ref{fig:vector_exp} shows the individual fits to the T2K and NO$\nu$A datasets in the vector ULDM models. The neutrino oscillation probabilities are computed by assuming the best-fit values that were previously obtained for the vector DM parameters with $L_e - L_\mu$ and $L_\mu - L_\tau$ charges. Under the hypothesis of vector DM, the neutrino oscillation parameter fits for T2K remain largely robust. For NO$\nu$A, while the $90\%$ CL allowed regions exhibit substantial overlap with the SM contours, a discernible shift is observed in the fitting results: the preferred value of $\sin^2\theta_{23}$ decreases and $\delta_{CP}$ shifts slightly toward the region favored by T2K. Despite these localized adjustments, the inclusion of DM does not yield a significant alleviation of the persistent tension between the T2K and NO$\nu$A datasets. These results are qualitatively consistent with those obtained in flavor-universal scenarios.

\section{Conclusions}
\label{sec:concl}

In the work, we have performed a comprehensive analysis of the ULDM signatures in long-baseline neutrino oscillations, utilizing the latest joint datasets from the T2K and NO$\nu$A experiments. We investigated the phenomenological implications of several ULDM coupling scenarios, including flavor-universal and flavor-general couplings in the scalar sector, as well as the $L_e - L_\mu$ and $L_\mu - L_\tau$ gauge symmetries in the vector sector. 

Crucially, our analysis explicitly incorporates the stochastic nature of the ULDM field. We introduce a Bayesian statistical treatment to account for the amplitude fluctuations determined by the ratio of the experimental exposure time to the DM coherence time. Our results show that in the low-mass regime ($m_\phi \lesssim 10^{-17}$ eV), where the stochastic effects are maximal, the constraints derived on coupling constants are relaxed by approximately one order of magnitude compared to the deterministic case, specifically by factors of $\sim 8.6$ for the scalar, $\sim 8.0$ for the $L_e - L_\mu$ and $\sim 8.5$ for the $L_\mu - L_\tau$ case, respectively. The mass range of $10^{-17}~\text{ eV} \lesssim m_\phi \lesssim 10^{-15}~\text{eV}$ serves as a transition away from the stochastic-dominated regime. Conversely, for $m_\phi \gtrsim 10^{-15}$~eV, the system enters a multi-sample regime ($N_{\rm DM} \gg 1$) where the stochastic fluctuations are effectively suppressed, recovering the deterministic limits. In addition, we also find that the choice of the lightest neutrino mass has little impact on the overall constraints, leading only to a marginal relaxation of the excluded regions.

The joint T2K and NO$\nu$A datasets yield the following substantial $3\sigma$ exclusion limits in the maximal fluctuation region ($m_\phi \lesssim 10^{-17}$ eV):
\begin{itemize}
    \item \textbf{Scalar (flavor-universal):} ${y_0}/{m_\phi} \lesssim 19.4 \, \text{eV}^{-1}$;
    \item \textbf{Vector (\boldmath $L_e - L_\mu$):} ${g}/{m_\phi} \lesssim 7.41 \times 10^{-10} \, \text{eV}^{-1}$;
    \item \textbf{Vector (\boldmath $L_\mu - L_\tau$):} ${g'}/{m_\phi} \lesssim 4.08 \times 10^{-10} \, \text{eV}^{-1}$.
\end{itemize}
Regarding the model preference, the identified best-fit points obtained from a global scan over the full mass range for each scenario read
\begin{itemize}
    \item \textbf{Scalar (flavor-universal):} $m_\phi \simeq 6.31 \times 10^{-13}$ eV, $y_0 \simeq 7.93 \times 10^{-13}$;
    \item \textbf{Vector (\boldmath $L_e - L_\mu$):} $m_\phi \simeq 1.58 \times 10^{-13}$ eV, $g \simeq 7.93 \times 10^{-24}$;
    \item \textbf{Vector (\boldmath $L_\mu - L_\tau$):} $m_\phi \simeq 3.98 \times 10^{-13}$ eV, $g' \simeq 7.93 \times 10^{-24}$.
\end{itemize}
Although the inclusion of ULDM couplings leads to slight shifts in the preferred oscillation parameters, the statistical improvement over the standard oscillation is marginal ($\Delta \chi^2 < 1$). Consequently, we do not find compelling evidence that the current ULDM models provide a viable resolution to the existing parameter tension between T2K and NO$\nu$A. {\hx  This limited numerical sensitivity could be qualitatively understood from the algebraic structures of the effective potentials. Unlike conventional mechanisms such as non-standard interactions~\cite{Denton:2020uda,Chatterjee:2024kbn} that introduce a new complex phase via off-diagonal elements, the vector ULDM potential is generally diagonal, whereas the scalar ULDM modifications are inherently entangled with the neutrino mass and the PMNS mixing matrix. Such structural constraints make it less straightforward to effectively provide the independent phase degrees of freedom required to reconcile the $\delta_{CP}$ discrepancy.}

Looking ahead, we anticipate that future high-precision facilities, such as DUNE and JUNO, will be instrumental in further clarifying the $\delta_{CP}$ landscape and providing more decisive probes of the ULDM parameter space.

\acknowledgments
This work is supported in part by the National Natural Science Foundation of China under Grant Nos.~12347105, 12475094, and 12135006, as well as the Fundamental Research Funds for the Central Universities (23xkjc017) in Sun Yat-sen University. SV is grateful to Tommy Ohlsson and Pedro Pasquini for fruitful discussions during this project.

\bibliography{references}

\end{document}